\documentclass[amsmath,amssymb]{revtex4}
\usepackage[parfill]{parskip}    
\usepackage{graphicx}
\usepackage{amssymb}
\usepackage{epstopdf}
\usepackage{color}
\usepackage{graphicx}
\usepackage{pdfpages}

\begin{document}

\title{Beyond classical anisotropy and a new look to relativistic stars: a gravitational decoupling approach}

\author{Gabriel Abell\'an {\footnote{gabriel.abellan@ciens.ucv.ve}} }
\address{Departamento de F\'isica, Facultad de Ciencias,
Universidad Central de Venezuela, A.P. 47270, Caracas 1041-A, Venezuela.}
\author{\'Angel Rinc\'on {\footnote{angel.rincon@pucv.cl}}}
\address{Instituto de F\'isica, Pontificia Universidad Cat\'olica de Valpara\'iso, Avenida Brasil 2950, Casilla 4059, Valpara\'iso, Chile.}
\author{Ernesto Fuenmayor {\footnote{ernesto.fuenmayor@ciens.ucv.ve}}}
\address{Departamento de F\'isica, Facultad de Ciencias,
Universidad Central de Venezuela, A.P. 47270, Caracas 1041-A, Venezuela.}
\author{Ernesto Contreras {\footnote{econtreras@usfq.edu.ec}}}
\address{Departamento de F\'isica, Colegio de Ciencias e Ingenier\'ia, Universidad San Francisco de Quito, Quito, Ecuador.} 

\begin{abstract}
In this article, we propose a physical condition
to extend interior isotropic solutions to anisotropic domains by gravitational decoupling in the framework of the Minimal Geometric Deformation approach.  In particular, it is found that by using an expression reminiscent of the classical anisotropy factor, we can close the decoupling system of equations and a new anisotropic solution can be found. As an example, we extend the well--known Tolman IV. 
\end{abstract}
\maketitle

\section{Introduction}\label{Intro}
General Relativity \cite{GR} (GR hereafter) is usually accepted as one of the most successful theories in the context of modern physics. This theory represents a mathematical model parameterized by a set of differential equations usually supplemented by some additional conditions which let us to obtain information regarding the physical world. Also, GR has been successfully applied to 
exterior solutions (i.e., black holes and cosmological solutions) as well as interior ones (i.e., relativistic compact stars). Precisely, the latter case has been widely analyzed considering first the simplest cases (i.e., perfect fluids) and second, taking more complicated, and therefore realistic situations (i.e., imperfect solutions). In this respect, it is essential to point out the seminal work of Tolman \cite{r82}, who introduced one of the most famous and useful perfect fluid solution found up to now. Along years, a great number of isotropic solutions have been considered (see \cite{Buchdahl1959a,Bondi1964a,Delgaty:1998uy,Panotopoulos:2019wsy,Panotopoulos:2019zxv} and references therein), but undoubtedly, the more interesting cases emerge when the fluid is indeed anisotropic (see also \cite{DeBenedictis:2005vp,Lopes:2019psm,Thirukkanesh:2008xc}). Such interest is well justified given that the physics inside compact (relativistic) stars is quite complicated and non-linear physics takes place. Thus, certainly could be more appropriate the study of anisotropic fluid to get insight about the physics in such extreme situations.
In the context of anisotropic solutions, the history have started largely around 30's. To be more precise, the idea of ``local anisotropy" in the context of GR was first noticed by G. Lemaitre \cite{Lemaitre_1933}. In that seminal paper, the compactness factor (i.e., the ratio between the total mass and the radio of the stellar distribution) was introduced.  
Although some progress was made, it was just after the paper of Bowers and Liang \cite{Bowers:1974} when the effect of anisotropic fluid was indeed taken seriously.
Taking as inspiration the aforementioned work, L. Herrera, G. Ruggeri and L. Witten studied the stability problem as well as the adiabatic index of anisotropic spheres \cite{1979ApJ...234.1094H}, and later, a general formulation to obtain exact anisotropic solution in relativistic stars was proposed \cite{Herrera:1981}.
Moreover, the evolution of some of the models treated in such papers are investigated in \cite{1982PhRvD..25.2527C}.
A decade after those works, a non-trivial feature in relativistic stars was also reported. Such is the case of the now well-known cracking of relativistic spheres \cite{Herrera:1992lwz} ``which results from the appearance of
total radial forces of different signs in different regions of the sphere, once the equilibrium configuration has been perturbed'' \cite{Herrera:1992lwz}.
A quite detailed classical review can be consulted in Ref. \cite{Herrera:1997plx} and an algorithm to find solutions can be found in \cite{Contreras:2019dsf}.
Currently, we can find a robust list of manuscripts where the physics of isotropic/anisotropic stars is investigated, see for example \cite{Maurya:2019sfm,Bhar:2019gwk,Singh:2019dki,Errehymy:2019ume}.\\

Local anisotropy has been widely studied. Besides the works already mentioned, we have the solutions found by Bayin \cite{bayin1982}, Herrera and Ponce de Leon \cite{herrera-leon1985a,herrera-leon1985b}, Florides \cite{florides1974}, Maartens and Maharaj \cite{maartens1990}, Melfo and Rago \cite{melfo1992} among others. There are a large number of processes that give rise to local anisotropies: highly dense systems \cite{collins1975}, exotic phase transitions during gravitational collapse \cite{hartle1975}. Other sources of anisotropy are found in viscocity \cite{kazanas1978,barreto1992}, collisionless systems \cite{michie1963,kent1982,cuddeford1991} as well as slow rotating systems \cite{kippenhahn2012}. In all the works previously mentioned, solutions have been found by building suitable models for the matter and solving Einstein's equations directly.\\

In this work, in order to deal with anisotropic solutions, we will implement
gravitational decoupling in the framework of
the so--called Minimal Geometric Deformation approach (MGD)\cite{Casadio:2015jva,Casadio:2015gea,Ovalle:2016pwp,Ovalle:2017fgl,Ovalle:2017wqi,Ovalle:2018umz,Ovalle:2018ans,r95,r97}. The MGD method was proposed in Randall-Sundrum brane-world context 
\cite{Ovalle:2009xk,Casadio:2012rf}
but recently has been applied to
decoupling gravitational sources in
classical GR \cite{Ovalle:2017fgl,Ovalle:2017wqi}. In the same spirit, the method has been applied to extend
standard relativistic compact objects \cite{Heras:2018cpz,Gabbanelli:2018bhs,Estrada:2018zbh,Morales:2018nmq,Morales:2018urp,MauryaTello-Ortiz2019,Tello-Ortiz:2019gcl,Torres:2019mee,isot,Abellan2} as well as black holes, wormholes  and even cosmology \cite{Rincon:2019jal,Gabbanelli:2019txr,Contreras:2018gzd,Cedeno:2019qkf,contrerasads}. Also, in dimensions different than four we can find well-known solutions (see \cite{Contreras:2018vph,Contreras:2019fbk,Contreras:2019iwm,Panotopoulos:2018law,Estrada:2018vrl} and references therein) and beyond standard GR \cite{Maurya:2019hds,Sharif:2019gkf, Sharif:2018khl,Sharif:2018brq,estradaLovelock}. 
In addition, MGD has been used to introduce non-trivial deviations to holographic entanglement entropy \cite{daRocha:2019pla}, and the quantum portrait of a black hole \cite{Fernandes-Silva:2019fez}.
A particularly powerful feature of the method is that it can handle easily problems involving anisotropy. To be more precise, in the context of MGD some isotropic model is considered as a sector of the total solution, so that it becomes in seed which is used to obtain an anisotropic solution of Einstein's equations. In this sense, the
main advantage of MGD is that it takes a known solution and extends it to more complicated and not trivial cases. Our purpose in this work will be to find 
alternative constraints to extend isotropic solutions to anisotropic domains.\\

The present work is organized as follow: after this brief introduction, we present the basic equations of GR as well as the MGD formalism. Then, in section III, we 
propose and alternative constraint to close the system in the framework of MGD. Next in section IV we briefly review the results related to the Tolman IV solution and then, in Sect VI, we show how our map works. Finally, in section VI we will briefly summarize and conclude the main features of the present article.
We adopt the metric signature, $(+, -, -, -)$, and
we work in geometric units where $c = G = 1$.

\section{Field equations and MGD}\label{sec:1}

In Schwarzschild like coordinates, a static and spherically symmetric geometry is described by the following line element
\begin{equation}\label{mgd01}
    ds^2 = e^{\nu}dt^2 - e^{\lambda} dr^2 - r^2(d\theta^2 + \sin^2{\!\theta}\,d\phi^2)\,, 
\end{equation}
where the metric functions, $\nu=\nu(r)$ and $\lambda=\lambda(r)$, depend only on the radial coordinate. We are interested in solving the Einstein's equations for an energy-momentum tensor $T_{\mu\nu}$ of the form
\begin{equation}\label{mgd02}
    T_{\mu\nu} = T^{(o)}_{\mu\nu} + \alpha\Theta_{\mu\nu}\;.
\end{equation}
The $T^{(o)}_{\mu\nu}$ term corresponds to a perfect fluid
\begin{equation}\label{mgd03}
    T_{\mu\nu}^{(o)} = (\rho^{(o)} + P_r^{(o)})u_{\mu} u_{\nu} - P_r^{(o)} g_{\mu\nu}\,, 
\end{equation}
and the $\Theta_{\mu\nu}$ term, describing any extra source, is coupled by means of the parameter $\alpha$. 
%
It is essential to point out that the additional term $\alpha \Theta_{\mu \nu}$ is not considered a perturbation, i.e., the coupling parameter $\alpha$ could indeed be larger than unity. Thus, such coupling is introduced in order to control the effect of the unknown anisotropic source.
%
\\

Now, given the parametrization (\ref{mgd01}) and the energy-momentum tensor (\ref{mgd02}), the Einstein's field equations\footnote{We shall use $\kappa=8\pi$ in explicit calculations},
\begin{equation}\label{mgd04}
    R_{\mu\nu} - \frac{1}{2}R g_{\mu\nu} = -\kappa T_{\mu\nu}\;,
\end{equation}
are written explicitly in components as follows
\begin{eqnarray}
    \kappa\! \left(\rho^{(o)} + \alpha\Theta^0_0\right) &=& \frac{1}{r^2} +
        e^{-\lambda}\!\left(\frac{\lambda'}{r} - \frac{1}{r^2}\right)\!,\label{mgd05}\\
    \kappa\! \left(P_r^{(o)} - \alpha\Theta^1_1\right) &=& -\frac{1}{r^2} +
        e^{-\lambda}\!\left(\frac{\nu'}{r} + \frac{1}{r^2}\right)\!,\label{mgd06}\\
    \kappa\! \left(P_r^{(o)} - \alpha\Theta^2_2\right) &=& \frac{1}{4}e^{-\lambda}\!\left(2\nu'' + {\nu'}^2 - \lambda'\nu' + 2\frac{\nu'-\lambda'}{r} \right)\!,\label{mgd07}
\end{eqnarray}
where the primes denote derivatives with respect to radial coordinate. The left hand side of these equations can be related with effective the quantities
\begin{eqnarray}
    \tilde{\rho} &=& \rho^{(o)} + \alpha\Theta_{0}^{0} \;,\label{mgd07a}\\
    \tilde{P}_{r} &=& P_r^{(o)}-\alpha\Theta_{1}^{1}   \;,\label{mgd07b}\\
    \tilde{P}_{\perp} &=& P_r^{(o)} -\alpha\Theta_{2}^{2} \;.\label{mgd07c}
\end{eqnarray}
Note that, as in general $\Theta^1_1 \neq \Theta^2_2$, we find that the system corresponds to an anisotropic fluid
by construction.  It is important to remark that since the Einstein tensor in (\ref{mgd04}) is divergence free, the total energy-momentum tensor (\ref{mgd02}) must satisfy the condition
\begin{equation}\label{mgd08}
    \nabla_\nu T^{\mu\nu} = 0\;,
\end{equation}
which can be interpreted as a conservation equation. If we calculate the conservation equation, the radial component is given by
\begin{eqnarray}\label{mgd09}
    P'^{(o)}_r + \frac{\nu'}{2}
    \bigl(\rho^{(o)} + P_r^{(o)}\bigl)
    - \alpha \Bigl[
    \Theta'^1_{1} 
     - \frac{\nu'}{2} (\Theta^0_0 - \Theta^1_1)
     - \frac{2}{r}(\Theta^2_2 - \Theta^1_1) \Bigl]
    = 0\, ,
\end{eqnarray}
from where it is straightforward that the $\Theta_{\mu \nu}$ tensor undoubtedly disturbs the energy configuration of the corresponding stellar object. Also, when the coupling constant $\alpha$ (value which parameterizes the new anisotropy) is taken to be zero, we recover the usual (isotropic) solution.
The last equation corresponds to the improved version of the well-known Tolman-Oppenheimer-Volkov (TOV) equation. At this level, it is essential to note that Eq. (\ref{mgd08}) is satisfied and, given that we  know that for a perfect fluid satisfies
$\nabla_\nu T^{(o)\mu\nu} = 0$, then the following condition must necessarily be fulfilled
\begin{equation}\label{mgd10}
    \nabla_\nu \Theta^{\mu\nu} = 0\;.
\end{equation}
The above conditions express the fact that there is no exchange of energy momentum between the perfect fluid and source $\Theta^{\mu\nu}$, the interaction is purely gravitational. A detailed discussion regarding how this method works can be found in \cite{Ovalle:2009xk, Casadio:2015gea, Ovalle:2017wqi, Heras:2018cpz} and references therein. So, we will circumvent technical details to focus on the underlying physics.\\

To find a complete solution to the Einstein's equations and, in order to fix the parameters involved, some exterior solution need to be considered.
Given that we are in presence of a four dimensional spacetime, without charge or angular momentum, then it is suitable to use the most canonical exterior solution. That is the reason why
we will use the Schwarzschild exterior solution
\begin{equation}\label{sol04a}
    ds^2 = \left(1 - \frac{2M}{r}\right)dt^2 - \left(1 - \frac{2M}{r}\right)^{-1} dr^2 \,-\, r^2 d\Omega^2\,,
\end{equation}
with $d\Omega^2 = d\theta^2 + \sin^2{\!\theta}\,d\phi^2$. Then, in order to match the two metrics, interior (\ref{mgd01}) and exterior (\ref{sol04a}) on the boundary surface $\Sigma$, we require
\begin{eqnarray}
    e^{\nu}\Big|_{\Sigma} &=& \left(1 - \frac{2M}{r}\right)\Bigg|_{\Sigma}\,,\label{mgd11a}\\
    e^{\lambda}\Big|_{\Sigma} &=& \left(1 - \frac{2M}{r}\right)^{-1}\Bigg|_{\Sigma}\,,\label{mgd11b}\\
     P_r(r) \Big|_{\Sigma} &=& \;0\,,\label{mgd11c}
\end{eqnarray}
which corresponds to the continuity of the first and second fundamental form across the surface $\Sigma$.\\

The MGD scheme prescribes that the metric coefficients are modified as
\begin{eqnarray}
    \nu\;\; &\longrightarrow &\;\; \nu\;,\label{mgd11} \\
    e^{-\lambda^{(o)}}\;\; &\longrightarrow &\;\; e^{-\lambda^{(o)}} + \alpha f\;,\label{mgd12}
\end{eqnarray}
where $f\equiv f(r)$ is the so-called deformation function and, in this case, depends on the radial coordinate only. Now,
when we replace (\ref{mgd11}) and (\ref{mgd12}) in Eqs. (\ref{mgd05}), (\ref{mgd06}), (\ref{mgd07})
, we are able to split the complete set of differential equations into two subset as follows.
%
%
\noindent The first one satisfies the Einstein's field equations for the perfect fluid, i.e.,
    \begin{eqnarray}
    \kappa\rho^{(o)} &=& \frac{1}{r^2} +
        e^{-\lambda^{(o)}}\!\left(\frac{\lambda'^{(o)}}{r} - \frac{1}{r^2}\right)\!,\label{mgd13}
        \\
    \kappa P_r^{(o)} &=& -\frac{1}{r^2} +
        e^{-\lambda^{(o)}}\!\left(\frac{\nu'}{r} + \frac{1}{r^2}\right)\!,\label{mgd14}
        \\
    \kappa P_r^{(o)} &=& \frac{1}{4}e^{-\lambda^{(o)}} \!\! \left( 2\nu'' + {\nu'}^2 
    - \lambda'^{(o)}\nu' + 2\frac{\nu'-\lambda'^{(o)}}{r} \right) \!,\label{mgd15}
\end{eqnarray}
with the conservation law
\begin{align}
\nabla_\nu T^{(o)\mu\nu} &= 0.
\end{align}

\noindent The second set corresponds to a system of equations for $f$, namely 
    \begin{eqnarray}
    \kappa\Theta^0_0 &=& -\frac{f}{r^2} -\frac{f'}{r}\,,\label{mgd16}\\
    \kappa \Theta^1_1 &=& -f
        \left(\frac{\nu'}{r} + \frac{1}{r^2}\right)\!,\label{mgd17}\\
    \kappa \Theta^2_2 &=& -\frac{f}{4} \left( 2\nu'' + 
    {\nu'}^2 + 2\frac{\nu'}{r}\right)
    -\frac{f'}{4} \left( \nu' + \frac{2}{r} \right)\!,\label{mgd18}
    \end{eqnarray}
with the elements of $\Theta_{\mu\nu}$ satisfying the equation
\begin{eqnarray}\label{consthe}
    \Theta'^1_{1} 
     - \frac{\nu'}{2} (\Theta^0_0 - \Theta^1_1)      - \frac{2}{r}(\Theta^2_2 - \Theta^1_1)=0. 
\end{eqnarray}
Notice that, although the above system looks similar to the canonical Einstein fields equations, the prefactor $1/r^2$ is absent in the right-hand side of Eqs. (\ref{mgd16}) and (\ref{mgd17}), reason why we can not identify the spherically symmetric Einstein field equations with radial metric component $f(r)$. Thus, given such discrepancy, the aforementioned system is called a ``quasi-Einstein system", as was introduced in \cite{Ovalle:2017fgl}. However, it worth noticing that
although the above system does not have the standard form of Einstein's equations, the conservation Eq. (\ref{consthe}) corresponds to the TOV equation for the decoupling sector. Indeed, (\ref{consthe}) is a linear combination of the Eqs. (\ref{mgd16}), (\ref{mgd17}) and (\ref{mgd18}).\\

\section{Alternative constraint on the decoupling sector}
In order to extend any isotropic solution
in the framework of MGD, extra information regarding the $\Theta$--sector, namely, Eqs. (\ref{mgd16}), (\ref{mgd17}) and (\ref{mgd18}), is required. Indeed, the system corresponds to  three equations with four unknown, $\{f,\Theta^{0}_{0},\Theta^{1}_{1},\Theta^{2}_{2}\}$. 
%
%
Among all the possibilities, the so called mimetic constraints are broadly used. In particular the mimetic constraint for the pressure which reads
\begin{eqnarray}
P^{(o)}_r=\Theta^{1}_{1}\,,    
\end{eqnarray}
leads to an algebraic equation that allows to 
find the decoupling function, $f$, analytically. Another possibility is the mimetic constraint for the density, 
\begin{eqnarray}
\rho^{(o)}=\Theta^{0}_{0}\,,
\end{eqnarray}
which yields a differential equation for $f$ that have been use to extend the well known Tolman IV solution. 
In this work, we propose an alternative constraint based on the relationship between the different sectors involved.
In particular, the effective quantities given above (see Eqs. \eqref{mgd07a}, \eqref{mgd07b} and \eqref{mgd07c}), suggest us to re-interpreted such equations in a more convenient way. First, recall that the well--known anisotropy factor is defined as
\begin{align}
    \Delta \equiv P_{\perp} - P_r\,.
\end{align}
where the set $\{P_{\perp}, P_r\}$ corresponds to the same fluid. This last, and simple, relation reveals that if the system is isotropic, i.e, $\Delta = 0$, or not, i.e. $\Delta \neq 0$. 
However, in this work we restrict our attention to a sort of pseudo anisotropy that naturally emerges. 
To be more precise, note that Eq. (\ref{mgd07c}) can be rewritten as
\begin{eqnarray}
\tilde{P}_{\perp} - P_{r}^{(o)}=-\alpha \Theta^{2}_{2}\,,   
\end{eqnarray}
which corresponds to a kind of anisotropy function given by the difference between the transverse pressure of the total solution and the pressure of the isotropic sector,
\begin{eqnarray}\label{Deltaeffec}
\tilde{\Delta}\equiv\tilde{P}_{\perp} - P_{r}^{(o)}\;,
\end{eqnarray}
which can be thought as a kind of interaction term accounting the effect of one fluid over the other one. In the next section we shall propose a particular functional form of constraint (\ref{Deltaeffec}) to extend the Tolman IV solution.

\section{Generating new exact solutions}\label{sec:4}

This section is devoted to implement a particular isotropic solution and then verify that we are able to obtain a reasonable anisotropic solution using the identification we just discussed above. To this end, we shall consider the well-known Tolman IV model as our seed solution which metric functions read 
\begin{eqnarray}
    e^{\nu} &=& B^2\left(1+\frac{r^2}{A^2}\right)\,,\label{sol03} \\
    e^{\lambda^{(o)}} &=& \frac{\left( 1+\frac{2r^2}{A^2} \right)}{\left( 1-\frac{r^2}{C^2} \right) \left( 1+\frac{r^2}{A^2} \right)}\,,\label{sol04}
\end{eqnarray}
that correspond to a solution of Einstein's equations sourced by 
\begin{eqnarray}
    \rho^{(o)}(r) &=& \frac{3A^4 + A^2(3C^2 + 7r^2) + 2r^2(C^2 + 3r^2)}
        {\kappa C^2 (A^2 + 2r^2)^2}\,,\;\;\label{sol05}\\
    P_r^{(o)}(r) &=& \frac{C^2 - A^2 - 3r^2}{\kappa C^2 (A^2 + 2r^2)}\,.\label{sol06}
\end{eqnarray}
The matching conditions (\ref{mgd11a}--\ref{mgd11c}) for this solution are given by the following expressions
\begin{eqnarray}
    1 - \frac{2M_o}{R} &=&  B^2\left(1 + \frac{R^2}{A^2} \right) , \label{sol07}\\
    1 - \frac{2M_o}{R} &=& \frac{\left( 1- \frac{R^2}{C^2} \right) \left( 1 + \frac{R^2}{A^2} \right)}{\left( 1 + \frac{2R^2}{A^2} \right)}  \,,\label{sol08}
    \\
    0 &=& \frac{C^2 - A^2 - 3R^2}{\kappa C^2(A^2 + 2R^2)}\,,\label{sol09}
\end{eqnarray}
and therefore we finally obtain 
\begin{eqnarray}
    \frac{A^2}{R^2} &=& \frac{R}{M_o} - 3\,,\label{sol10}\\
    B^2 &=& 1 - \frac{3M_o}{R}\,,\label{sol11}\\
    \frac{C^2}{R^2} &=& \frac{R}{M_o}\,\label{sol12}.
\end{eqnarray}

Now, using our ansatz (\ref{Deltaeffec}), we can proceed to apply the MGD method by introducing the anisotropy condition in Eq. (\ref{mgd18}) which yield
\begin{equation}\label{sol19}
    \kappa \tilde{\Delta} = \frac{1}{4}f\!\left(2\nu''+\nu'^2+2\frac{\nu'}{r}\right) 
        +\frac{1}{4}f'\!\left(\nu'+\frac{2}{r}\right)\,.
\end{equation}
The next step consists in to 
provide a particular form for $\tilde{\Delta}$ which leads to a differential equation for the decoupling function $f$. We shall explore this in what follows.


\subsection{A toy model} 
Let us propose as a pseudo anisotropy the following function
\begin{equation}\label{sol20}
    \tilde{\Delta} = \beta r^2\,.
\end{equation}
Note that, apart from its simplicity,
there is nothing special about this choice of $\tilde{\Delta}$. From the metric coefficient (\ref{sol03}) we find that
\begin{equation}\label{sol21}
    \nu' = \frac{2r}{A^2 + r^2}\;, \hspace{.6cm}
    \nu'' = \frac{2(A^2 - r^2)}{(A^2 + r^2)^2}\;.
\end{equation}
Next, after replacing (\ref{sol21}) in  (\ref{sol20}) and solving the differential equation we find 
\begin{equation}\label{sol23}
    f = \frac{A^2 + r^2}{(A^2+2r^2)^{3/2}}c_1 - \frac{\kappa\beta}{15}(A^2-3r^2)(A^2+r^2)\,,
\end{equation}
where $c_1$ is an arbitrary integration constant to be obtained later.\\

Using the prescription (\ref{mgd11}) and (\ref{mgd12}), we find that the modified metric coefficients are
\begin{eqnarray}
      e^{\nu} &=& B^2\left(1+\frac{r^2}{A^2}\right)\,,\label{sol26} \\
      e^{-\lambda^{(1)}} &=& 
      e^{-\lambda^{(0)}}
      +\,\alpha \Bigg[ \frac{\left(1 + \frac{r^2}{A^2}\right)}{A\left(1 + \frac{2r^2}{A^2}\right)^{3/2}}\,c_1 
       - \frac{\kappa\beta}{15}A^4\left(1 - \frac{3r^2}{A^2}\right)\left(1 + \frac{r^2}{A^2}\right)\Bigg]\;.\label{sol27}
\end{eqnarray}
The superindex $(1)$ indicates that we are considering the first application of the method. At this level, we can fix the value for the integration constant $c_1$ noticing that we want to interpret the metric coefficient as $e^{-\lambda^{(1)}} = 1 - 2m_1/r$, so that by consistency, the radial metric coefficient should satisfied \cite{Bondi1964a}
\begin{equation}\label{sol28}
    e^{-\lambda^{(1)}} \;\; \longrightarrow \;\; 1 \hspace{.6cm} \mbox{when} \hspace{.6cm} r\to 0\;,
\end{equation}
and therefore, using these condition we find that the integration constant that appear in (\ref{sol23}) is fixed to
\begin{equation}\label{sol29}
    c_1 = \frac{\kappa\beta}{15} A^5\;.
\end{equation}
With this value for $c_1$ the new metric coefficient in MGD is
\begin{eqnarray}
      e^{-\lambda^{(1)}} &=& e^{-\lambda^{(o)}}
      +\,\alpha \frac{\kappa\beta}{15}A^4 \Bigg[ \frac{\left(1 + \frac{r^2}{A^2}\right)}{\left(1 + \frac{2r^2}{A^2}\right)^{3/2}}
      - \phantom{\frac{\phantom{\kappa}}{\beta}}\!\!\!
      \left(1 - \frac{3r^2}{A^2}\right)\left(1 + \frac{r^2}{A^2}\right)\Bigg] \!. \label{sol30}
\end{eqnarray}
Now that we have found the new metric coefficients, we can calculate the new solution.

\subsection{New effective quantities}
As discussed above, once we have found the solution for $f$ we can insert this in the pseudo Einstein system and then find expressions for $\Theta^0_0$ and $\Theta^1_1$. Doing so, we obtain
\begin{eqnarray}
    \Theta^0_0\; &=&\; -\frac{\beta}{15}\!\!\left[ 6A^2 + 15r^2 - \frac{A^4}{r^2} 
    + \frac{A^9 - A^7 r^2}{r^2(A^2+2r^2)^{5/2}} \right]\!,\;\;\;\;\label{sol30a} \\
    \Theta^1_1\; &=&\; \frac{\beta}{15r^2}(A^2+3r^2)\!\!\left[ A^2 - 3r^2 
    - \frac{A^5}{(A^2+2r^2)^{3/2}} \right]\!.\;\;\;\;\label{sol30b}
\end{eqnarray}
Using these results and the Eqs. (\ref{mgd07a}), (\ref{mgd07b}) and (\ref{mgd07c}), we obtain the new effective quantities corresponding to the anisotropic fluid
%
\begin{align}
    &\tilde{\rho} = \rho^{(o)} - 
    \frac{2}{5}A^2 \alpha\beta
    \Bigg[
    1 + \frac{5r^2}{2A^2} -
    \frac{A^2}{ 6r^2} + 
    \frac{1 -  \frac{r^2}{A^2}}{\frac{6 r^2}{A^2} (1 + \frac{2r^2}{A^2})^{\frac{5}{2}}}
    \Bigg] ,
    \\
    &\tilde{P}_r = P_r^{(o)} - \frac{\alpha \beta A^4}{15 r^2}
    \Bigg(
    1 + \frac{3r^2}{A^2}
    \Bigg)\!
    \Bigg[
    1 - \frac{3r^2}{A^2} - \frac{1}{(1 + \frac{2r^2}{A^2})^{\frac{3}{2}}}
    \Bigg] ,
    \\
    &\tilde{P}_{\perp} =  P_r^{(o)} + \beta r^2\,.
\end{align}
\noindent An anisotropic contribution appears in this solution, this is evident from the fact that $\tilde{P}_r(r) \neq \tilde{P}_\perp(r)$.

\subsection{Matching conditions} 
%
%
In the following, we establish the appropriated matching conditions for the modified problem. Using Eqs. (\ref{sol26}), (\ref{sol27}), we match the metric coefficients with Schwarzschild exterior solution (\ref{sol04a}) and also establish the condition over the radial pressure $\tilde{P}_r(R) = 0$. The new matching conditions are
\begin{eqnarray}
    1 - \frac{2M_1}{R} &=& B^2\left(1+\frac{R^2}{A^2}\right) \!,\label{sol31}\\
    1 - \frac{2M_1}{R} &=& \frac{\left( 1-\frac{R^2}{C^2} \right) \left( 1+\frac{R^2}{A^2} \right)}{\left( 1+\frac{2R^2}{A^2} \right)}
    +\,\alpha \frac{\kappa\beta}{15} A^4
    \left[ \frac{\left(1 + 
    \frac{R^2}{A^2}\right)}{\left(1 + \frac{2R^2}{A^2}\right)^{3/2}} 
    -\!
    \left(1 - \frac{3R^2}{A^2}\right)\left(1 + \frac{R^2}{A^2}\right) \right]  \!,\label{sol32}\\
    0 \;&=&\; \frac{C^2 - A^2 - 3R^2}{C^2(A^2 + 2R^2)} 
    - \frac{\alpha \beta A^4}{15 R^2} \Bigg( 1 + \frac{3R^2}{A^2} \Bigg) 
    \left[\, 1 - \frac{3R^2}{A^2}
      - \frac{1}{\left(1+\frac{2R^2}{A^2}\right)^{3/2}} \right] \!.\label{sol33}
\end{eqnarray}
Note that these conditions define a new mass $M_1$ for the modified problem. Now, with the help of (\ref{sol08}) and the pressure condition (\ref{sol33}) we find that
\begin{eqnarray}
    B^2\left(1+\frac{R^2}{A^2}\right) &=&  1 - \frac{2M_o}{R}
    +\,\alpha A^4\frac{\kappa\beta}{15}\! \left[ \frac{(1 + \frac{R^2}{A^2})}{(1+\frac{2R^2}{A^2})^{3/2}}
     - \Big(1-\frac{3R^2}{A^2}\Big)\Big(1+\frac{R^2}{A^2}\Big) \right]  \!,\label{sol34}\\
    \frac{2M_1}{R} &=& \frac{2M_o}{R}
    -\,\alpha A^4\frac{\kappa\beta}{15}\! \left[ 
    \frac{(1 + \frac{R^2}{A^2})}{(1+\frac{2R^2}{A^2})^{3/2}}
    - \Big(1-\frac{3R^2}{A^2}\Big)\Big(1+\frac{R^2}{A^2}\Big) \right]  \!.\label{sol35}
\end{eqnarray}
Notice that the condition $\tilde{P}_r(R)=0$ allows to obtain a relation between the constant $C$ and the rest of free parameters. 
Also, it is essential to point out that these new matching conditions are consistent with the old ones when we take the limit $\alpha\to 0$. \\

\begin{figure*}[ht]
\centering
\includegraphics[width=0.32\textwidth]{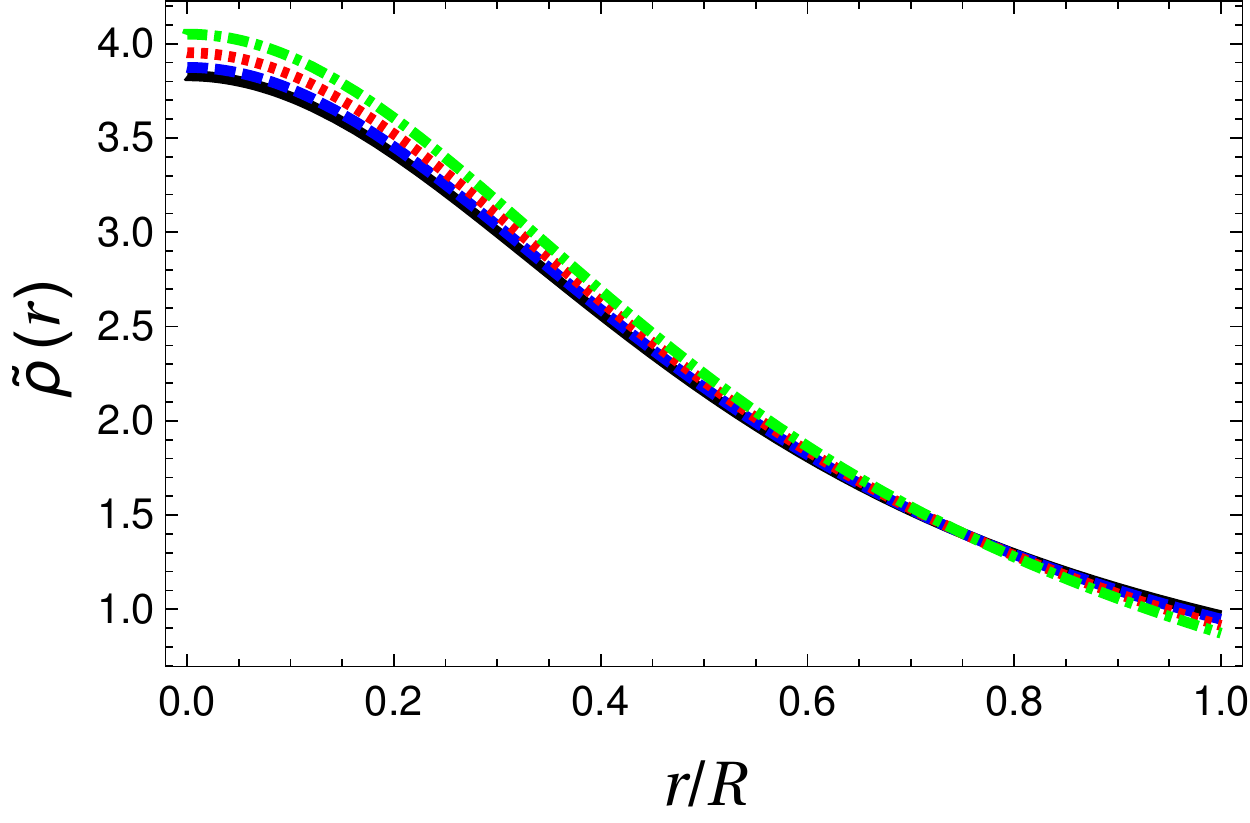}             \
\includegraphics[width=0.32\textwidth]{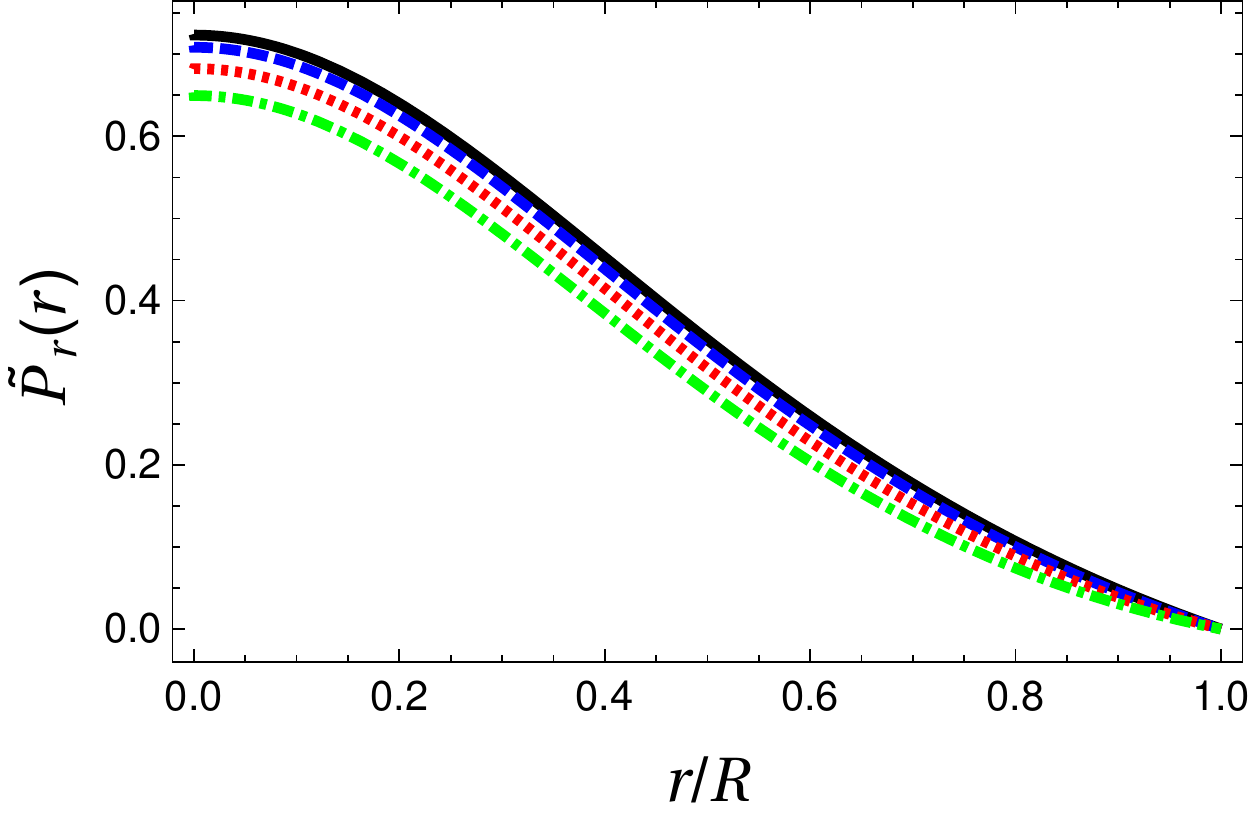}     \
\includegraphics[width=0.32\textwidth]{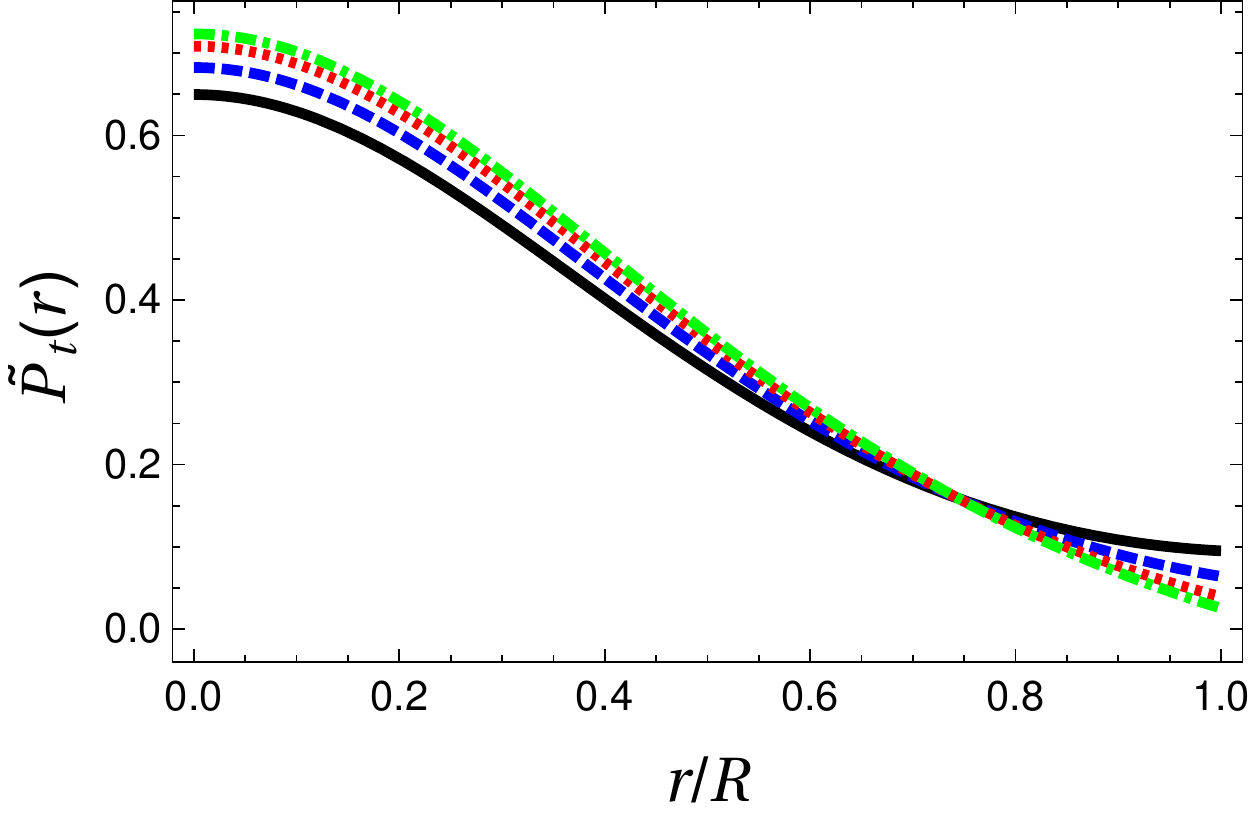} \

\medskip

\includegraphics[width=0.32\textwidth]{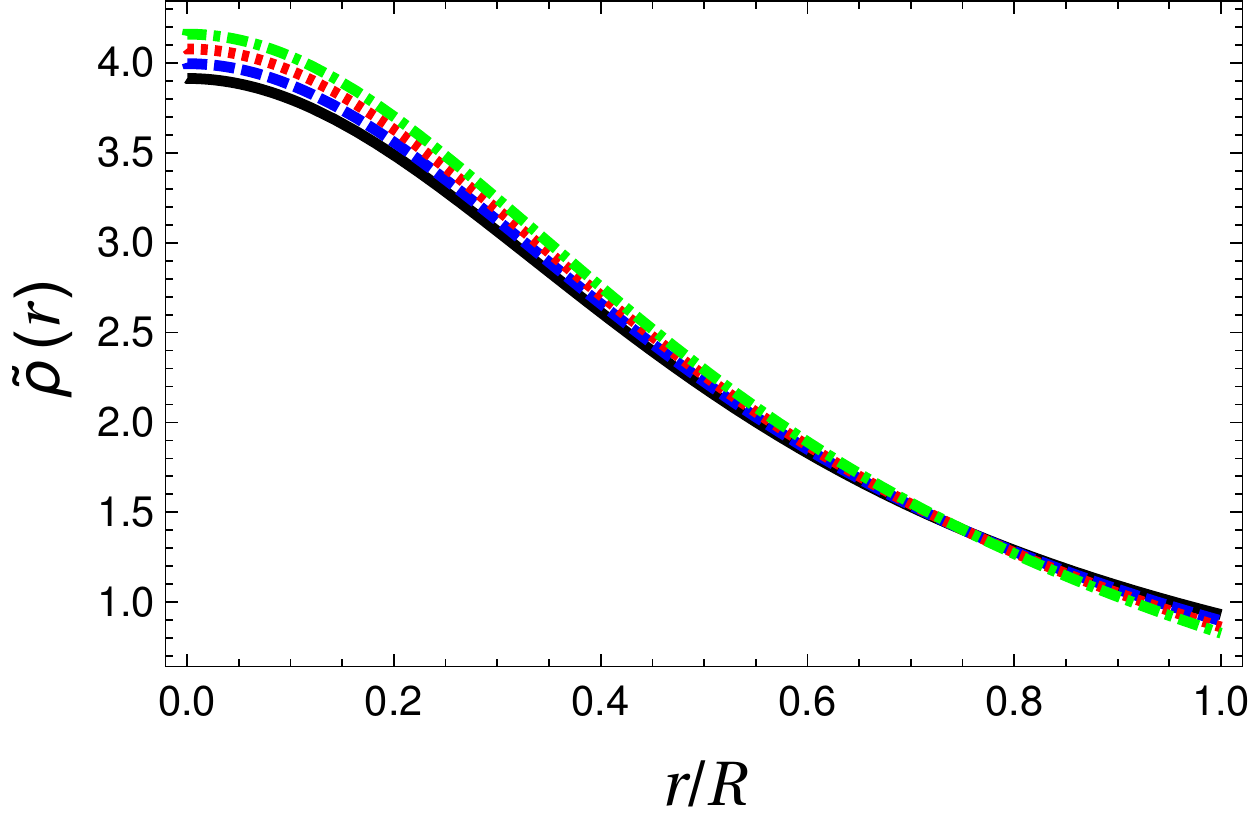}             \
\includegraphics[width=0.32\textwidth]{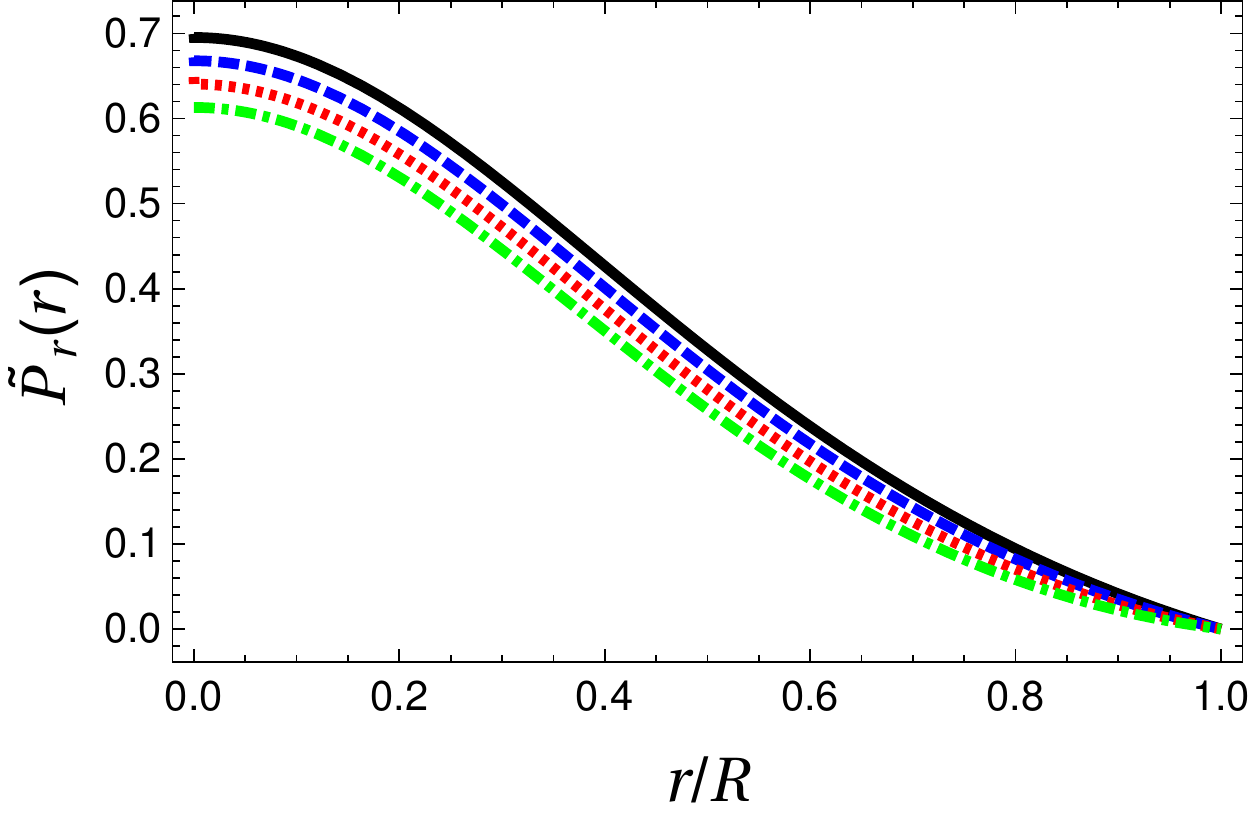}     \
\includegraphics[width=0.32\textwidth]{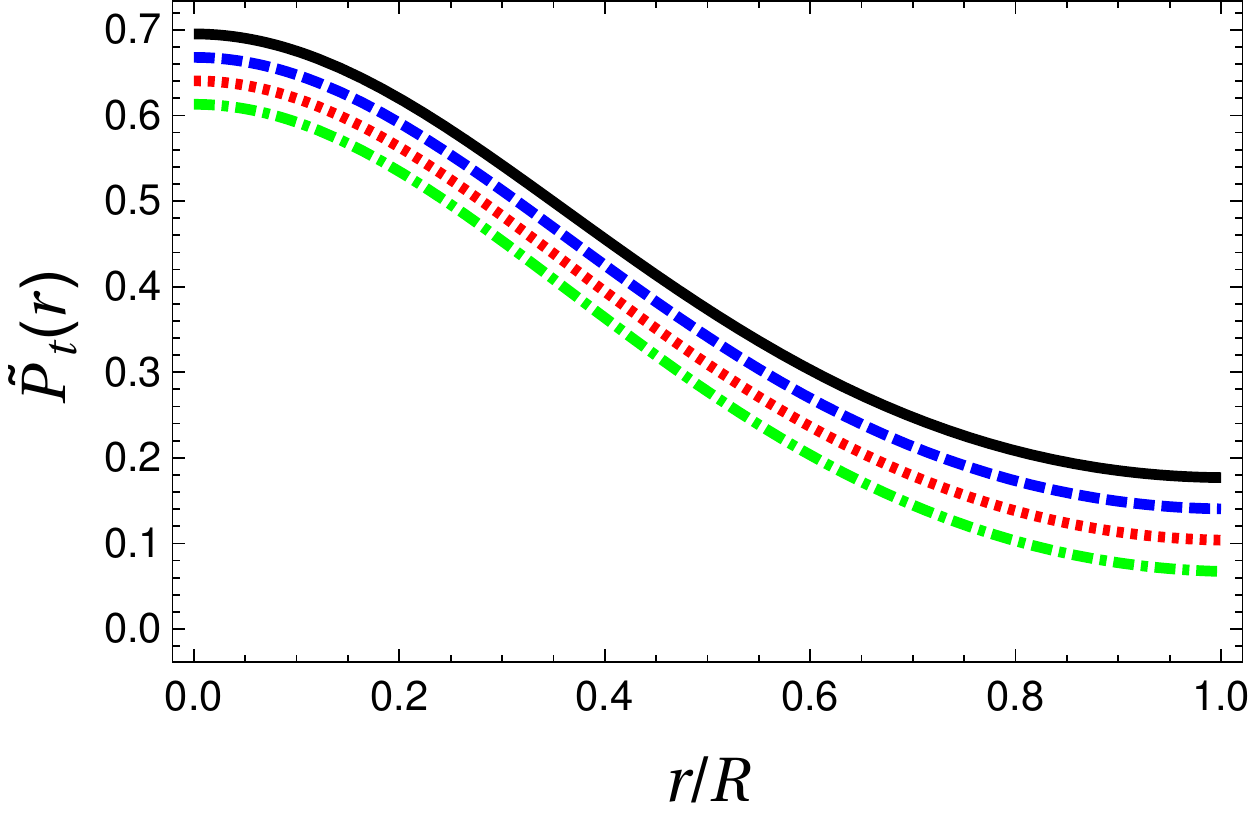} \
\caption{
{\bf Parameter values for the first row:} 
$A=1, R=1, \kappa=1$, $\alpha=1$ with 
i)   $\beta = 0.2288$ (solid black line), 
ii)  $\beta = 0.2512$ (dashed blue line),
iii) $\beta = 0.2525$ (dotted red line),
iv)  $\beta = 0.2537$ (dot-dashed green line).
{\bf Parameter values for the second row:} 
$A=1, R=1, \kappa=1$, $\beta=0.25$ with 
i)   $\alpha = 0.50$ (solid black line), 
ii)  $\alpha = 0.75$ (dashed blue line),
iii) $\alpha = 1.00$ (dotted red line),
iv)  $\alpha = 1.25$ (dot-dashed green line).
{\bf LEFT:} Effective density versus radial normalized coordinate.
{\bf MIDDLE:} Effective radial pressure versus radial normalized coordinate.
{\bf RIGHT:} Effective tangential pressure versus radial normalized coordinate.
}
\label{fig:Tolman_IV}
\end{figure*}

At this level, it is essential to point out
that a physically acceptable solution have densities and pressures positively defined. Also, they have a maximum at the center of the object and decrease monotonically towards the surface.
In Fig. \ref{fig:Tolman_IV} we observe: 
i) the density profile $\rho(r)$ as a function of the normalized radial coordinate (left panel). We notice that physical behaviour is adequate, i.e., starting from a maximum value at the center, the density decreases monotonically as expected. Note also that the isotropic density is lower at the center and higher on the boundary of the stellar distribution, respect to the anisotropic counterpart.
ii) The radial pressure profile versus the normalized radial coordinated (middle panel). This starts from certain maximum value $\tilde{P}_r(0) \equiv \tilde{P}_{\text{max}}$ and decreases going to zero when $r=R$. The anisotropic cases are always lower than the isotropic case, further all figures decrease to zero when $r=R$, even the isotropic case. Thus, the inclusion of anisotropies in the Tolman IV solution does not introduce any non-physical feature. Finally, 
iii) the tangential pressure profile versus the normalized radial coordinate (right panel). This last quantity decreases as it is expected for interior solutions. We also observe that, albeit we have take advantage of a sort of pseudo anisotropy factor, the correspondent tangential pressure profile is still physically well-defined.

\section{Concluding remarks}\label{final}
In this work we have proposed an alternative constraint to close the decoupling sector induced by the minimal geometric deformation method to extend isotropic solutions. To be more precise, we introduced a pseudo--anisotropy which can be thought as term that encodes the interaction between two fluids. As an example we used the well--known Tolman IV solution as a seed and a particular representation for 
the pseudo anisotropy. Then, after a careful analysis of the parameters involved, we showed that the system fulfills the requirement of an acceptable physical solution.\\

Before concluding this work, we would like to point out that if we consider the Eqs. (\ref{mgd07b}) and (\ref{mgd07c}), we can motivate the appearance of an anisotropy factor using components of the tensor $\Theta_{\mu\nu}$. So we propose that
\begin{equation}\label{conn01a}
    \Delta = P_\perp - P_r = -(\Theta^2_2 - \Theta^1_1)\,,
\end{equation}
This anisotropy $\Delta$ is completely analogue to those used in conventional approaches.
Furthermore, using equations (\ref{mgd17}) and (\ref{mgd18}), we get the following equation for the deformation function
\begin{eqnarray}\label{conn02a}
    \frac{f'}{4}\left( \frac{2}{r} + \nu' \right) + \frac{f}{4} \left(
    2\nu'' + \nu'^2 - 2\frac{\nu'}{r} - \frac{4}{r^2} \right) = \kappa \Delta\,.\;\;
\end{eqnarray}
Note that if we know the anisotropy factor, it is possible to solve for $f$ because $\nu$ is the metric coefficient of the isotropic seed solution. Once we find $f$, we can reconstruct the other components of $\Theta_{\mu\nu}$ using the pseudo Einstein equations and therefore have all the effective quantities.\\

Another variant consists of study the pseudo--TOV equation (\ref{mgd10}) which explicitly reads as
\begin{equation}\label{conn03a}
        \Theta'^1_{1} 
     - \frac{\nu'}{2} (\Theta^0_0 - \Theta^1_1) - \frac{2}{r}(\Theta^2_2 - \Theta^1_1) = 0\;.
\end{equation}
So, using this equation and expression (\ref{conn01a}), together with equations (\ref{mgd16}) and (\ref{mgd17}) we find that the deformation function must satisfy
\begin{eqnarray}\label{conn04a}
    f'\left( \frac{1}{r^2} + \frac{3 \nu'}{2r} \right)\! - f \left( \frac{2}{r^3} + \frac{\nu'}{r^2}
    + \frac{\nu'^2}{2r} - \frac{\nu''}{r} \right) = -\frac{2\Delta}{r} \,.\;\;\;\;
\end{eqnarray}
Again, once we find $f$, we can reconstruct the other components of $\Theta_{\mu\nu}$ using pseudo Einstein equations. Alternatives (\ref{conn02a}) and (\ref{conn04a}) can be used to extend isotropic to anisotropic solutions via minimal geometric deformation methodology. However, these aspects are beyond the scope of this paper and we leave them as future works.\\

\section*{Acknowlegements}

The author A. R. acknowledges DI-VRIEA for financial support through Proyecto Postdoctorado 2019 VRIEA-PUCV.

\end{document}